\documentclass[review]{elsarticle}

\usepackage{fixltx2e}
\usepackage{hyperref}
\usepackage{ dsfont }
\usepackage{amsmath }
\usepackage{ bbold }
\usepackage{ amssymb }
\usepackage{epstopdf}

\usepackage{setspace}
\begin{document}

\begin{frontmatter}

\title{Dark solitons, modulation instability and breathers in a chain of weakly non-linear oscillators with cyclic symmetry}

\author[myfirstaddress]{F. Fontanela\textsuperscript{*}}
\ead{filipe.fontanela15@imperial.ac.uk}
\author[mysecondaddress]{A. Grolet}
\author[myfirstaddress]{L. Salles}
\author[mythirdaddress]{A. Chabchoub}
\author[myfirstaddress,myfourthaddress]{N. Hoffmann}

\address[myfirstaddress]{Department of Mechanical Engineering, Imperial College London, Exhibition Road, \\ SW7 2AZ London, UK}
\address[mysecondaddress]{Department of Mechanical Engineering, Arts et M\'etiers ParisTech, 8 Boulevard Louis XIV, 59000 Lille, France}
\address[mythirdaddress]{Department of Mechanical Engineering, Aalto University, P.O. Box 11000,\\ FI-00076 Aalto, Finland}
\address[myfourthaddress]{Department of Mechanical Engineering, Hamburg University of Technology, 21073 Hamburg, Germany}

\begin{abstract}
In the aerospace industry the trend for light-weight structures and the resulting complex dynamic behaviours currently challenge vibration engineers. In many cases, these light-weight structures deviate from linear behaviour, and complex nonlinear phenomena can be expected. We consider a cyclically symmetric system of coupled weakly nonlinear undamped oscillators that could be considered a minimal model for different cyclic and symmetric aerospace structures experiencing large deformations. The focus is on localised vibrations that arise from wave envelope modulation of travelling waves. For the defocussing parameter range of the approximative nonlinear evolution equation, we show the possible existence of dark solitons and discuss their characteristics. For the focussing parameter range, we characterise modulation instability and illustrate corresponding nonlinear breather dynamics. Furthermore, we show that for stronger nonlinearity or randomness in initial conditions, transient breather-type dynamics and decay into bright solitons
appear. The findings suggest that significant vibration localisation may arise due to mechanisms of nonlinear modulation dynamics.
\end{abstract}


\end{frontmatter}

\section{Introduction}

Localisation of vibrations has received considerable attention from the structural dynamics engineering community over the last three decades (see e. g.  papers \cite{Hodges1982,Hodges1983,Cornell1989,Vakakis1996,Bendiksen2000,Castanier2002,Georgiades2009,Mbaye2012} and references therein). In a linear framework, localisation may arise due to imperfections in the manufacturing process that result in a slightly disordered and inhomogeneous system. The localisation of vibrations has particular relevance in the aerospace industry, where e.g. bladed-disks of turbo-machines, reflectors, and antennas are usually composed of ideally periodic structures. Linear localisation due to disorder in general was first observed in solid state physics \cite{Anderson1958}, and this community usually refers to the phenomenon as Anderson Localisation. In the aerospace and turbo-machinery community, localisation due to imperfections is usually referred to as a mistuning \cite{Whitehead1966,Petrov2003}, when viewed from a spectral framework. Mistuning plays a significant role in system dynamics due to its importance in mechanical vibrations, fatigue, and even aerodynamics \cite{Krack2016}.  

In the case of turbo-machinery applications, more and more so-called blisks are used in current aero-engine designs. A blisk is a blade-integrated disk, i.e. a rotating component in which the traditional separation between a disk and attached blades is overcome enabling the whole structure to be formed monolithically. Blisks thus do not have any internal mechanical joints, with their corresponding friction mechanisms, and they are usually considered undamped structures \cite{Bartels2007}. Moreover, the dynamical response of blisks under operation is often thought to be beyond the traditional range of applicability for linear models due to the effect of geometric nonlinearity for large vibration amplitudes. 

A number of publications (e.g. references \cite{Vakakis1996,Georgiades2009,Kerschen2009,Grolet2011,Grolet2012,Starosvetsky2013,Grolet2015,Papangelo2017,Smirnov2017}) have already shown that perfectly symmetric or perfectly periodic structures may localise vibrations in the nonlinear regime due to: (i) the dependence of mode shapes on displacement amplitude, which is usually referred to as non-similar modes; and (ii) through bifurcations of main normal mode branches. The role and effects of such localised solutions are still under study within the vibration engineering community, and also the relevance of geometric nonlinearities when manufacturing variability plays any role is a question of active research \cite{Capiez-Lernout2015}. 

The emergence of travelling wave vibration states in aerospace structures has already been studied extensively, e.g. due to Coriolis effect \cite{Ruffini2015} or aeroelastic excitation \cite{Krack2016}. However, there is also a vast body of knowledge and literature on the weakly nonlinear dynamics of nonlinear travelling waves from the physics community that does not seem to be reflected widely in the present context \cite{Flach1998,Flach2008}. Especially in the optics community the study of modulation instability has lead to substantial progress in understanding and influencing the systems under study \cite{Dudley2014}. The aim of this research is to investigate the mechanisms of vibration localisation that may arise due to modulation of travelling waves. We therefore study the stability of plane wave vibration states and the subsequent nonlinear evolution here with a view to structural dynamics engineering. 

We use a highly idealised model for which we numerically study localised solutions inspired by insights from the Nonlinear Schrödinger Equation (NLSE). We find that depending on the wavelength of the travelling wave under study, the system becomes self-modulating, in the focussing parameter range, or the system can self-demodulate, in the defocussing parameter range \cite{Remoissenet1994}. In the self-modulating regime, the travelling nonlinear waves are linearly unstable against long-wavelength perturbations, in accordance with the general theory of modulation instability. Only in the self-demodulating regime are travelling waves linearly stable against long modulations. With a view to vibration localisation, the stable regime allows so-called dark solitons to arise, where spatially confined parts of the system do not vibrate, while the rest of the system is filled with a travelling wave. The unstable regime, as has already been reported in Ref. \cite{Grolet2016}, shows bright solitonic structures, but also a complex nonlinear evolution of modulation dynamics. Depending on system parameters and initial conditions, either breather-like vibration states emerge, complex dynamics involving breathers or soliton chaos. In all cases our numerical simulations suggest that very strong vibration localisations may arise when the weakly nonlinear system properties are taken into account. 

The paper is organized as follows. Section \ref{Sec:Model} introduces the analysed physical model and presents the physical framework required to deal with modulated nonlinear travelling waves. Solutions based on numerical integration are presented in Sec. \ref{Sec:NonSol}. Dark solitons are introduced for defocussing range of parameters while, for the focussing parameter range, modulation instability and corresponding nonlinear breathers emergence  are discussed in detail. Subsequently, Sec. \ref{Sec:RandSol} investigates the evolution of random initial conditions in the defocusing and focusing ranges. Finally, Sec. \ref{Sec:Con} summarises the conclusions and suggests directions for future investigations.

\section{The model and solution methods \label{Sec:Model} } 

\begin{figure}[h]
	\centering
	\includegraphics[trim=.5cm 4.5cm .5cm 4.cm, clip=true, angle=0, scale=0.35]{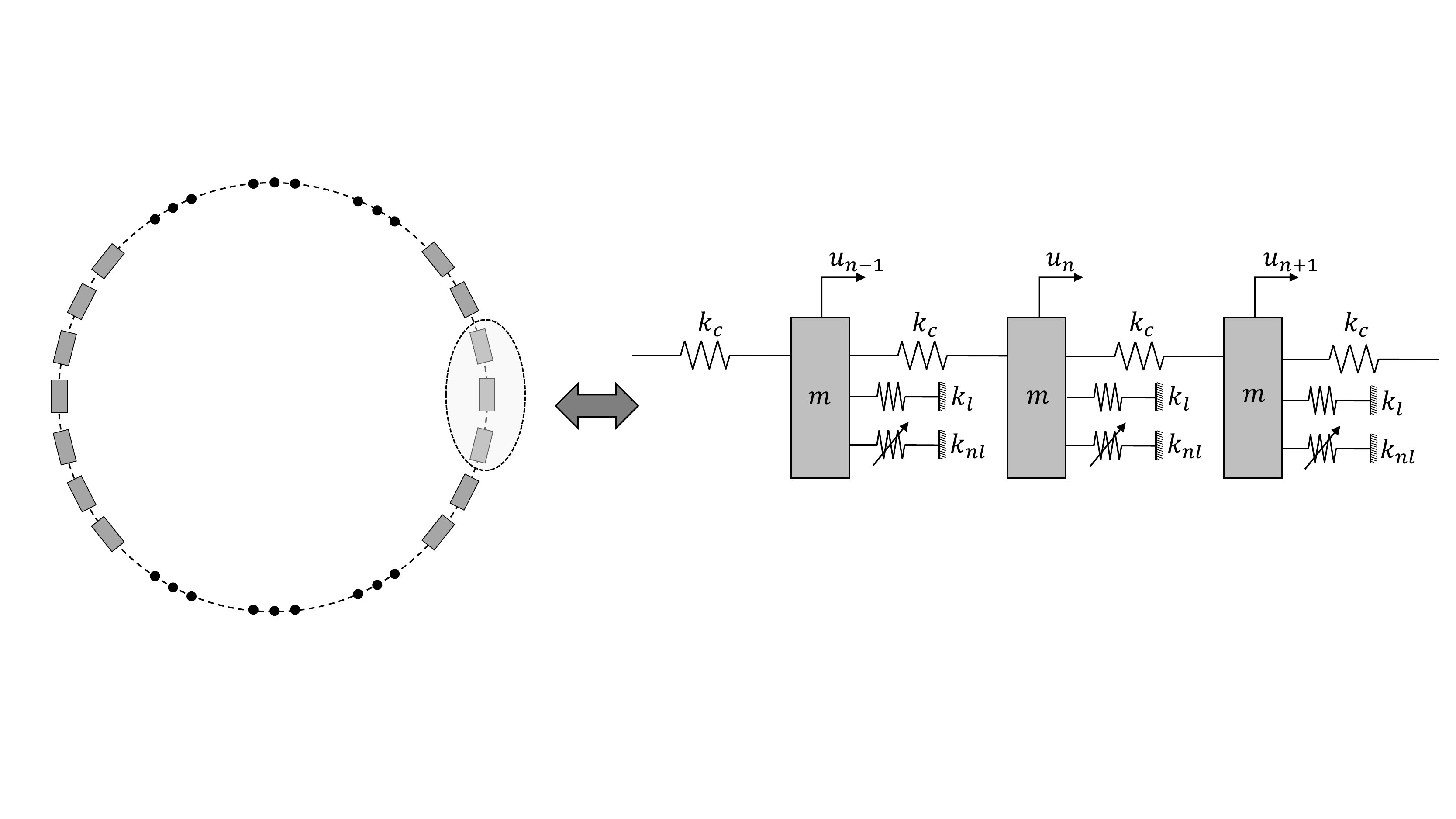}
	\caption{Mechanical system studied. On the left hand side, an illustration of the full system, while the right hand side displays any three neighbouring degrees of freedom.}
	\label{Fig:PhySys}
\end{figure}

The system under study consists of $N_s$ identical masses $m$ cyclically connected through linear springs with constant $k_c$. Each mass is also connected to the ground by a linear spring $k_l$, and a nonlinear one $k_{nl}$ of cubic behaviour. The corresponding configuration is depicted in Fig. \ref{Fig:PhySys}.
Physical systems similar to the one presented in Fig. \ref{Fig:PhySys} may be understood as a minimal model of different aerospace structures, such as space reflectors \cite{Cornell1989}, disk antennas \cite{Bendiksen1987} and bladed-disks of aero-engines \cite{Castanier2002,Georgiades2009}. In the case of bladed-disk vibrations, the model presented in Fig. (\ref{Fig:PhySys}) is obtained when the Von Karman theory is applied to a system of $N_s$ beams or plates that are cyclically coupled by massless springs, taking into account only the first term of a Rayleigh-Ritz approximation for the displacement field \cite{Grolet2011,Grolet2012}. In this case, nonlinearity arises due to large deformations and a positive (or negative) value of $k_{nl}$ represents a hardening (or softening) effect. 

The mathematical model for the $n$-th displacement $u_n$ in Fig. \ref{Fig:PhySys} is described by
\begin{equation}
m\ddot{u}_n + (k_l+2k_c)u_n - k_c(u_{n-1} + u_{n+1}) + k_{nl}u_n^3 = 0.
\end{equation}
The latter equation can be rewritten for convenience as 
\begin{equation}
\ddot{u}_n + \omega_0^2 u_n + \xi u_n^3 - \omega_c^2(u_{n-1}+u_{n+1} - 2u_n) = 0,
\label{Eq:EqMot}
\end{equation}
where $\omega_0^2=k_l/m$, $\omega_c^2=k_c/m$, and $\xi=k_{nl}/m$. 

In the linear regime ($\xi\equiv0$), Eq. (\ref{Eq:EqMot}) has exact plane harmonic wave solutions \cite{Grolet2011,Grolet2012,Grolet2016}
\begin{equation}
u_n(k)=U_k\exp\{i[k(n-1)a - \omega_k t]\} \ + c.c.,
\label{Eq:LinearPW}
\end{equation}
where $U_k$ is the amplitude, $a=2\pi/N_s$ is the sector parameter, $k$ is the wave number, $\omega_k=\sqrt{\omega_0^2 + 2\omega_c^2 ( 1 - \cos(ka))}$ is the dispersion relation, $i$ is the imaginary unity, and $c.c.$ states the complex-conjugate of the first expression. It should be noted that Eq. (\ref{Eq:LinearPW}) is usually valid when $k$ is an integer number and, consequently, $u_1=u_{N_s+1}$, due to cyclic symmetry, fulfil the cyclic system. For nonlinear vibrations we focus on modulations of the constant amplitude wave, using a continuous envelope function $\Psi$, such that
\begin{equation}
u_n(t)=\epsilon \Psi(X,T)\exp\{i[k(n-1)a - \omega_k t]\} \ + c.c,
\label{Eq:PropSol}
\end{equation}
where $X=\epsilon x = \epsilon (n-1)a$ and $T=\epsilon t$ denote scaled space and time variables, respectively. In Eq. (\ref{Eq:PropSol}), the term $\epsilon$ has been introduced as a small parameter such that for asymptotically weak nonlinearity a NLSE can be derived. It has been demonstrated that the slowly varying weakly nonlinear regime of Eq. (\ref{Eq:EqMot}) is governed by the following NLSE \cite{Remoissenet1994,Dauxois2006}
\begin{equation}
i \dfrac{\partial \Psi}{\partial \tau} + P \dfrac{\partial^2 \Psi^2}{\partial \eta^2} + Q |\Psi|^2\Psi = 0.
\label{Eq:NLS}
\end{equation}
In Eq. (\ref{Eq:NLS}), $\tau=\epsilon^2 t$ is a slowly varying time scale, $\eta=X - c_g T$ is the spatial variable in a frame moving with the group velocity $c_g=\dfrac{d\omega_k}{dk}$, $P=\dfrac{1}{2}\dfrac{d^2\omega_k}{dk^2}$ is the dispersion parameter, and $Q=-\dfrac{3\xi}{2\omega_k}$ is the term that accounts for nonlinearities \cite{Grolet2016}. Equation (\ref{Eq:NLS}) only considers first-order nonlinear terms and is based on the continuous approximation. In practical terms, the approach works well when the modulation $\Psi$ varies much slower in time and space than the carrier wave period $2\pi/w_k$ and the lattice parameter 2$\pi/N_s$, respectively. 

\begin{figure}[h]
	\centering
	\includegraphics[trim=4.25cm 0.75cm 4.cm 0.75cm, clip=true, angle=0, scale=0.275]{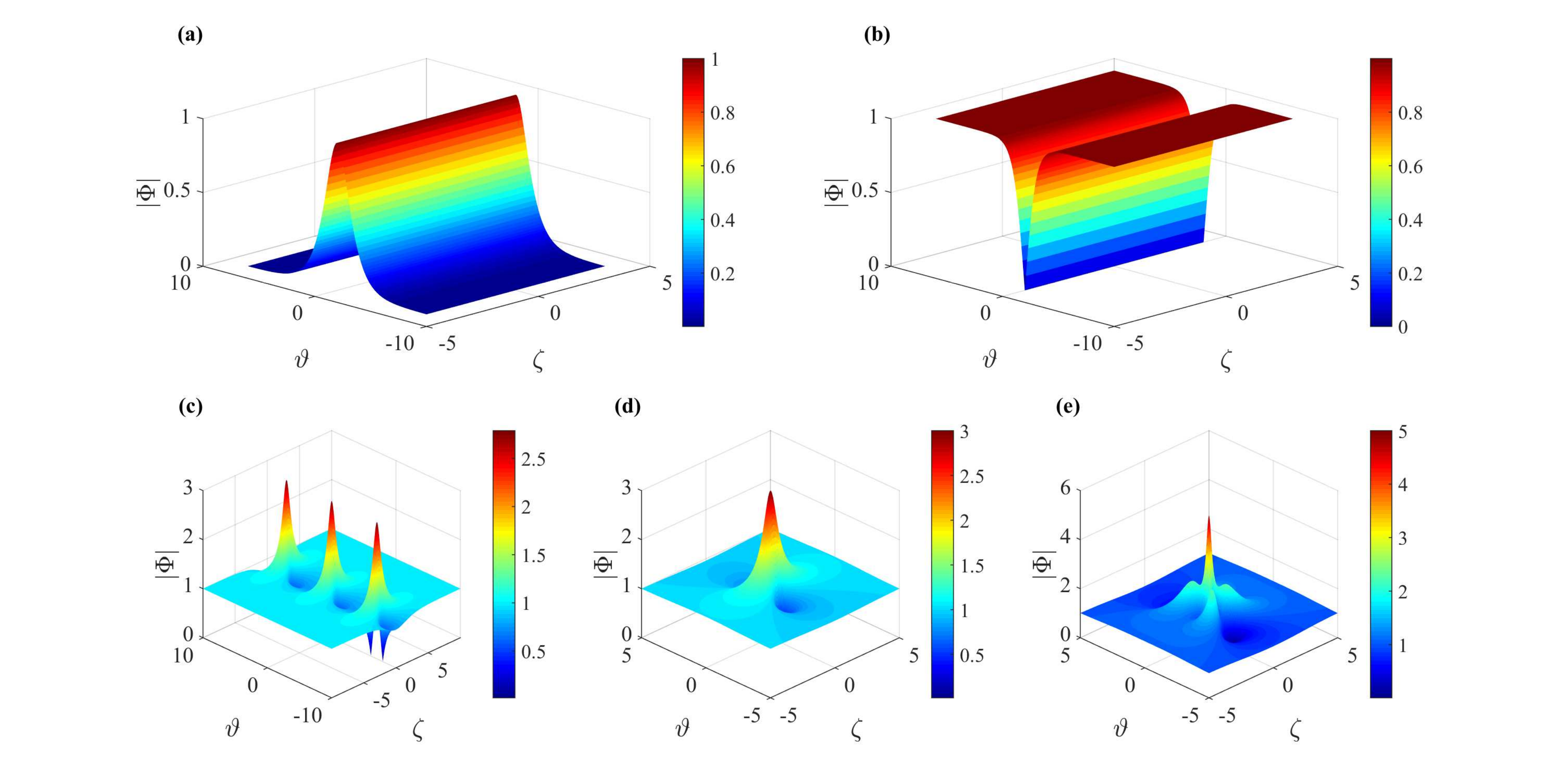}
	\caption{Localised solutions. Panels (a) and (b) illustrate the bright and dark solitons, respectively. Panels (c) and (d) display the Akhmediev breather, for $\alpha=0.4$, and the Peregrine solution. Panel (e) depicts the second-order rational expression, also known as the Akhmediev-Peregrine solution. All values are obtained following the scales presented in Sec. \ref{Sec:NonSol} and assuming an unit background amplitude.}
	\label{Fig:Appendix}
\end{figure}

Through the paper we use solutions of the continuous NLSE heuristically, and then study these solutions and their evolution in the context of the discrete physical system. Therefore, values from the continuous expression, obtained from the solution of Eq. (\ref{Eq:NLS}), are discretely sampled to generate initial conditions for Eq. (\ref{Eq:EqMot}). In this paper we are interested in the fundamental localised solitons and the breathers illustrated in Fig. \ref{Fig:Appendix}. Panels (a) and (b) present stationary bright and dark solitons, respectively. They consist of spatial localisation patterns that move around the system keeping a constant profile, with bright solitons arising for the focusing parameters, while dark solitons emerge for the defocusing parameters. Panels (c), (d) and (e) display Akhmediev, Peregrine and Akhmediev-Peregrine solutions, respectively, each of which arises for the focusing parameters only. They represent spatio-temporal modulation at which vibrations do not localise only in space but also in time. These features are investigated more fully in the next sections. 

\section{Nonlinear solutions} \label{Sec:NonSol}

\begin{figure}[h]
	\centering
	\includegraphics[trim=4.25cm 0.75cm 4.45cm 0.15cm, clip=true, angle=0, scale=0.27]{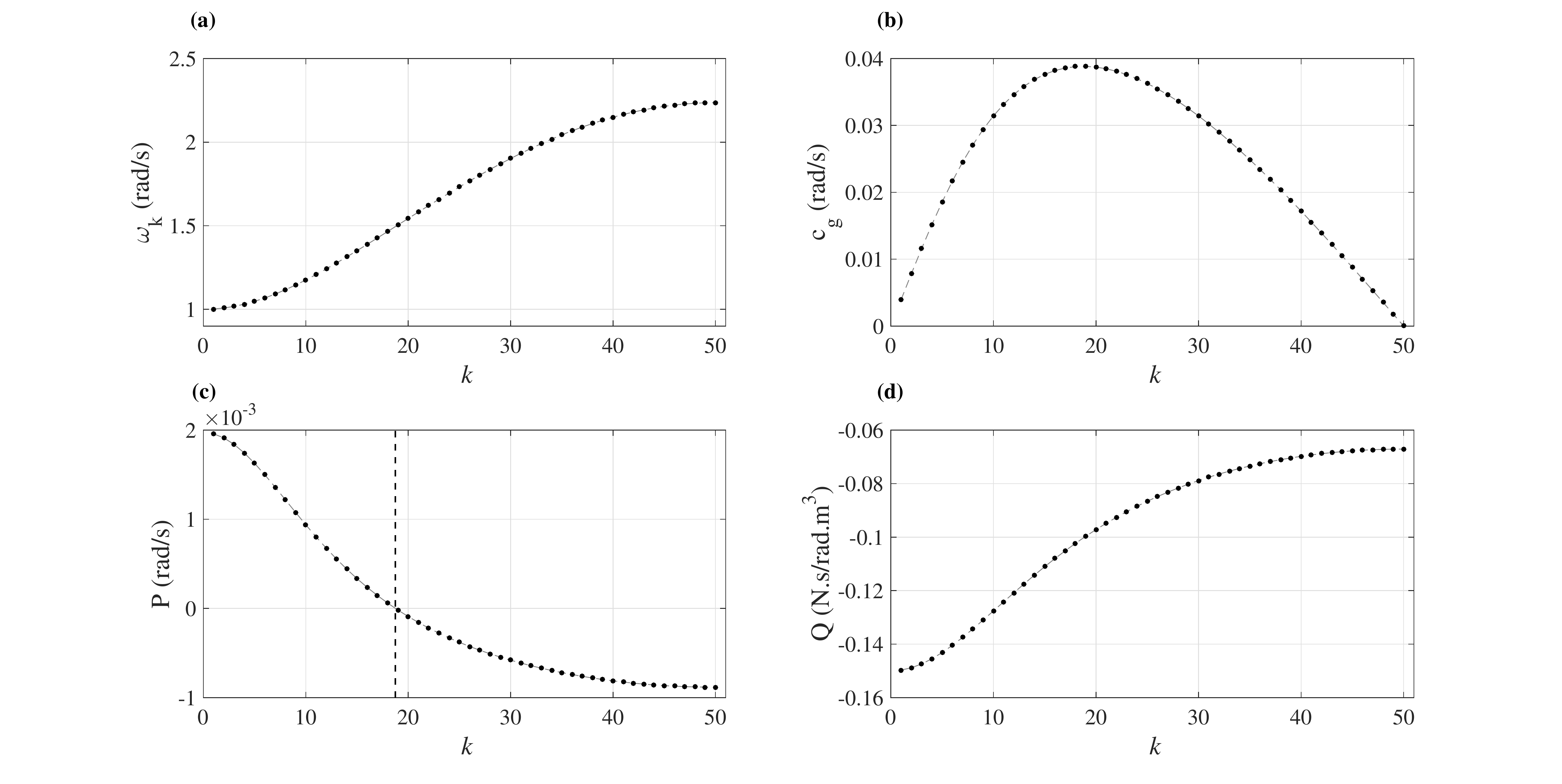}
	\caption{The system parameters as a function of the wave number $k$: (a) linear natural frequency $\omega_k$; (b) group velocity $c_g$; (c) dispersion parameter $P$, where the vertical dashed line states the transition from defocusing ($P>0$) to focusing ($P<0$) range; (d) nonlinear coefficient $Q$.}
	\label{Fig:SysPar}
\end{figure}

The system depicted in Fig. \ref{Fig:PhySys} is numerically investigated by assuming $N_s$=100, $\omega_0^2=1$ s$^{-2}$, $\omega_c^2=1$ s$^{-2}$, and $\xi=0.1$ m$^{-2}$s$^{-2}$. Figure \ref{Fig:SysPar} displays the values of the linear natural frequency $\omega_k$, the group velocity $c_g$, the dispersion parameter $P$, and the nonlinear coefficient $Q$, both as functions of the wave number $k$.
One should note that the physical system is cyclically symmetric, which means that the initial conditions also need to be chosen accordingly, which is usually the case when $k$ is an integer number. The values of $k$ are bounded between [1,$N_s$], but only values between [1,$N_s/2$] are investigated, since the system is cyclic and all the values in Fig. \ref{Fig:SysPar} are symmetric with respect to $k$=50. In addition, from Fig. \ref{Fig:SysPar}, a transition from positive values of $P$ to negative ones between $k=18$ and $k=19$ is observed. This transition delineates the boundary between the defocusing range, when $PQ<0$, and the focusing one, when $PQ>0$. All subsequent simulations are obtained through numerical integration of Eq. (\ref{Eq:EqMot}) by using a standard Runge-Kutta method, and initial conditions are extracted from analytic solutions calculated from the NLSE. It should be noted that, usually, NLSE reported results are obtained by considering an infinite line as its spatial domain rather than a finite and cyclic system. However, we only study modulations that are sufficiently localised to fulfil the cyclic system and thus boundary effects do not need to be considered here, as will be seen also from the results, as described below.

\subsection{Dark solitons}

It has already been shown in Ref. \cite{Grolet2016} that the proposed system has bright (or envelope) solitons in the focusing range. In this subsection we complement this finding by demonstrating the existence of dark (or hole) solitons in the defocusing parameter range. Dark solitons are well known from many other physical systems, such as fibre optics \cite{Emplit1987}, plasmas \cite{Shukla2006}, or water waves \cite{Chabchoub2013}, and this discussion for completeness describes how dark solitons look in the present context. We also would like to note that we here focus on so-called black solitons only \cite{Kivshar1994}.

\begin{figure}[h]
	\centering
	\includegraphics[trim=1.25cm 0.cm 1.25cm 0.25cm, clip=true, angle=0, scale=0.265]{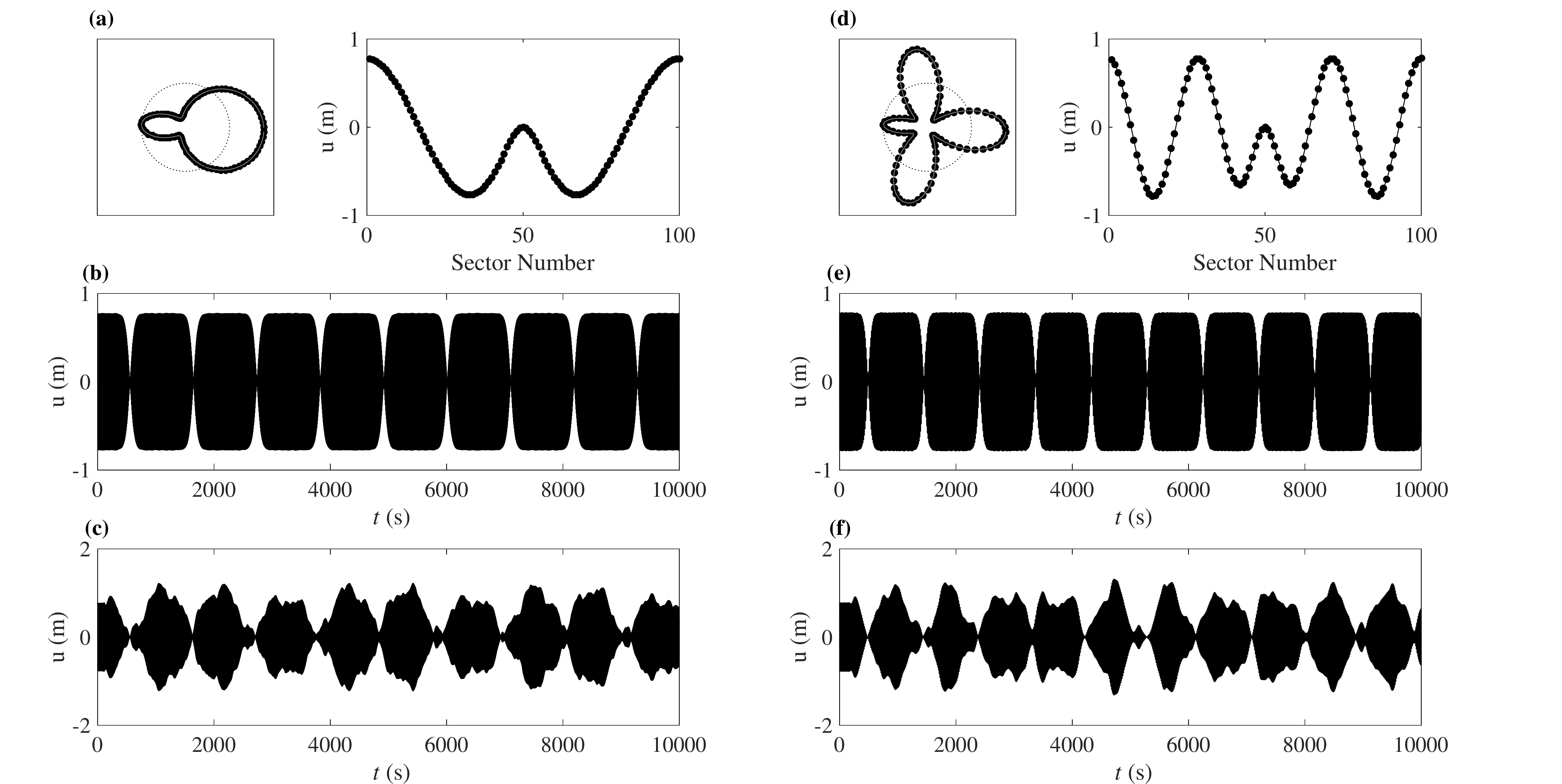}
	\caption{Dark solitons obtained by assuming an amplitude $\dfrac{V}{\sqrt{-Q}}\sim 0.39 \ m$, while the carrier wave is $k=1.5$, for the left side plots, and $k=3.5$, for the right side plots. Panel (a) illustrates the initial conditions for the system displacement (the unfolded graph is also plotted), Panel (b) depicts the displacement of a specific mass obtained from the nonlinear model, while Panel (c) displays the same quantity obtained with the equivalent linear model. Panels (d), (e), and (f) show the same quantities obtained by assuming $k=3.5$.}
	\label{Fig:HoleSolitons}
\end{figure}

Firstly, the NLSE in Eq. (\ref{Eq:NLS}) is investigated in a dimensionless form such that
\begin{equation}
i \dfrac{\partial \Phi}{\partial \zeta} +  \dfrac{\partial^2 \Phi}{\partial \vartheta^2} - 2|\Phi|^2\Phi=0,
\label{Eq:DimNLS1}
\end{equation} 
where $\vartheta=\dfrac{\eta}{\sqrt{2P}}$, $\zeta=\dfrac{\tau}{2}$, and $\Psi(\tau,\eta)=\dfrac{\Phi_D(\zeta,\vartheta)}{\sqrt{-Q}}$. It is well-known that Eq. (\ref{Eq:DimNLS1}) has dark soliton solution such that \cite{Bekki1985,Peregrine1983}
\begin{equation}
\Phi_{D}= i V \tanh (V\vartheta) \exp \{-2iV^2\zeta\},
\label{Eq:DSol}
\end{equation}
where $V$ is the amplitude of the background wave. The expression in Eq. (\ref{Eq:DSol}) represents a wave depression that keeps its stationary shape while it moves around the system with the group velocity, see Panel (a) of Fig. \ref{Fig:Appendix}. Initial conditions obtained from Eq. (\ref{Eq:DSol}) have a characteristic phase shift of $\pi$ centred in the wave train, see Panels (a) and (d) of Fig. \ref{Fig:HoleSolitons}. This means that an integer wave number, $k$, which originally fitted the cyclic system, would not fulfil the dark soliton initial conditions. This issue is solved by assuming a carrier wave number that does not fit the initial conditions due to a discontinuity of $\pi$ in phase, but due to the phase shift induced by the dark soliton solution does fit the cyclic system. Figure  \ref{Fig:HoleSolitons} displays the obtained numerical results. In this case, nonlinearity and dispersion are perfectly balanced and the hole in the initial conditions keeps a constant shape while it moves around the system with almost the value of the group velocity. For comparison, note that if the nonlinear terms in the system are ignored, the hole modulation spreads out when it propagates around the cyclic system due to dispersion only.  

\subsection{Modulation instability}

\begin{figure}[h]
	\centering
	\includegraphics[trim=10.5cm 0.75cm 9.75cm 0.5cm, clip=true, angle=0, scale=0.34]{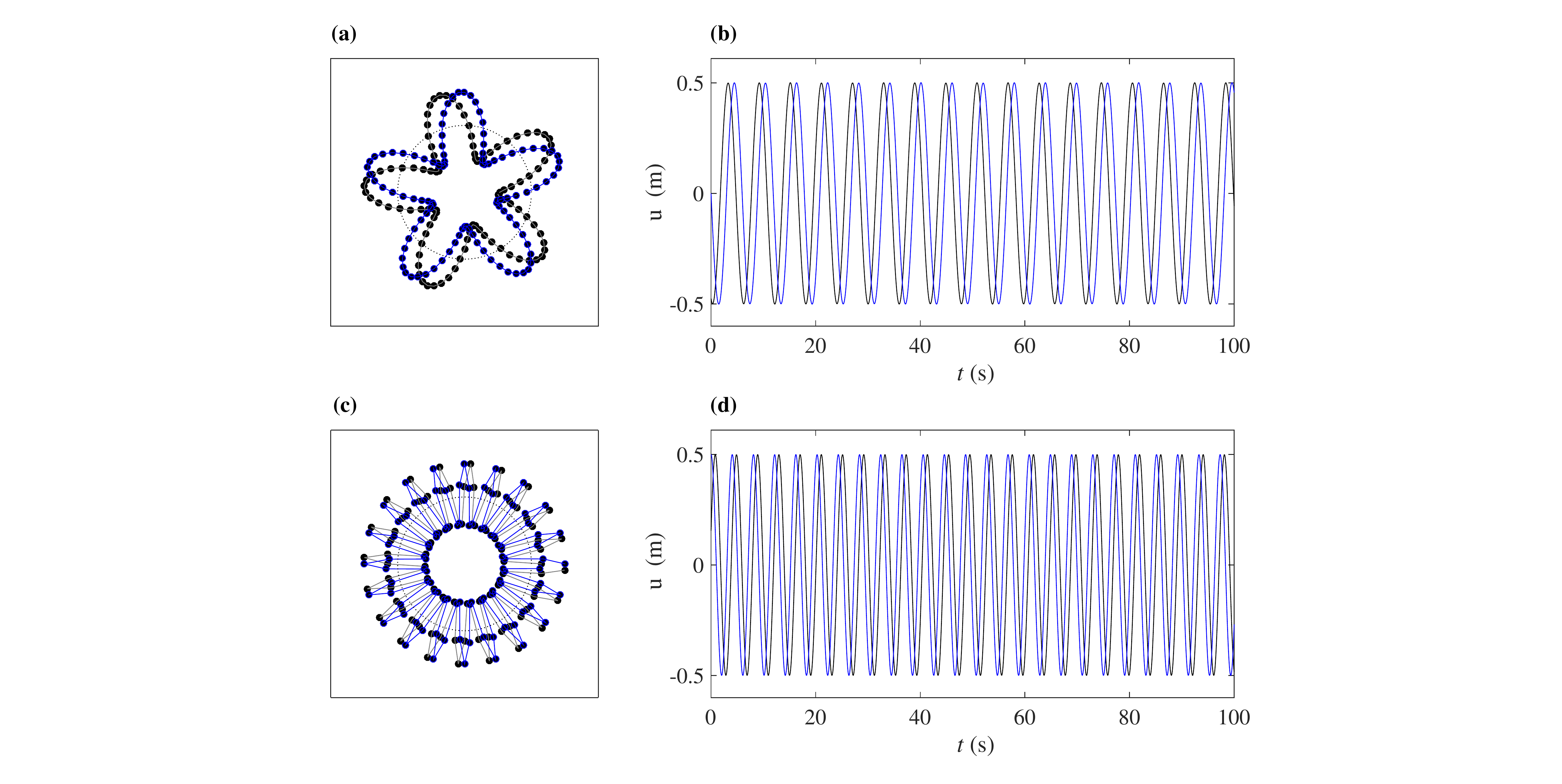}
	\caption{Plane wave solutions calculated by assuming the amplitude $V$=0.25 $m$ for two different wave numbers: $k=5$ and $k=20$. Panel (a) displays the spatial configuration for two different time values, while Panel (b) shows the displacement of two different degrees of freedom for $k=5$. Panel (c) and (d) illustrate the same quantities for $k=20$.}
	\label{Fig:PlaneWaves} 
\end{figure}

After investigating bright and dark solitons, it is useful to consider the linear stability of the plane wave solution itself. In the context of the NLSE this investigation can be started as follows. The nonlinear plane wave $\Psi$ is written as \cite{Remoissenet1994}
\begin{equation}
\Psi=V \exp\{i\theta(\tau)\} ,
\label{Eq:PlaneWave}
\end{equation}
where $V$ is a constant amplitude and $\theta(\tau)$ is an unknown phase function. After substituting Eq. (\ref{Eq:PlaneWave}) into Eq. (\ref{Eq:NLS}), the plane wave solution is written as 
\begin{equation}
\Psi=V \exp\{iQV^2 \tau\}.
\label{Eq:PWsol1}
\end{equation}
The calculated expression illustrates that the physical system oscillates with the linear travelling wave frequency $\omega_k$ corrected by a term proportional to $V^2$ due to the inherent nonlinearity. Figure \ref{Fig:PlaneWaves} displays the numerical results obtained by
assuming $V=0.25$ for two different wave numbers: (1) $k=5$ simulates the system in the defocusing range; and (2) $k=20$ computes results in the focusing range. The plotted curves show that a plane wave solution with constant and finite amplitude moves around the cyclic structure. 

Indeed, it is well-known from studying the NLSE that plane waves are not always linearly stable \cite{Remoissenet1994,Dauxois2006}. Therefore, the stability of plane waves due to small perturbations is investigated by assuming the following NLSE solution \cite{Remoissenet1994}:
\begin{equation}
\Psi(\eta,\tau)=(V + \epsilon V_{pert}(\eta,\tau)) \exp\{i[\theta(\tau) + \epsilon\theta_{pert}(\tau,\eta)]\} + c.c.
\label{Eq:ModInst1}
\end{equation}
In Eq. (\ref{Eq:ModInst1}), the term $\epsilon$ is a small parameter that controls the perturbation amplitude $V_{pert}(\eta,\tau)$ and phase $\theta_{pert}(\tau,\eta)$. After inserting Eq. (\ref{Eq:ModInst1}) into Eq. (\ref{Eq:NLS}), the standard plane wave solution is obtained at zeroth order. To order $\epsilon$ a set of coupled linear partial differential equations such that
\begin{eqnarray}
V \dfrac{\partial \theta_{pert}}{\partial \tau} - 2QV^2V_{pert} - P \dfrac{\partial^2 V_{pert}}{\partial \eta^2} &=& 0  \label{Eq:ModInst2},\\
\dfrac{\partial V_{pert}}{\partial \tau} + PV\dfrac{\partial^2 \theta_{pert}}{\partial \eta^2}&=&0 \label{Eq:ModInst3},
\end{eqnarray}
has to be solved. Equations (\ref{Eq:ModInst2}) and (\ref{Eq:ModInst3}) admit a pair of harmonic solutions \cite{Remoissenet1994}
\begin{eqnarray}
V_{pert}(\eta,\tau)=V_p \exp\{i(K\eta - \Omega \tau)\} + c.c, \label{Eq:ModInst4}\\
\theta_{pert}(\eta,\tau)=\theta_p \exp\{i(K\eta- \Omega \tau)\} + c.c \label{Eq:ModInst5},
\end{eqnarray}
where $V_p$ and $\theta_p$ denote two constants, while $K$ and $\Omega$ state the wave number and its corresponding frequency. After inserting Eqs. \ref{Eq:ModInst4} and \ref{Eq:ModInst5} into Eqs. \ref{Eq:ModInst2} and \ref{Eq:ModInst3}, the dispersion relation 
\begin{equation}
\Omega^2 = (K^2 - 2 \dfrac{Q}{P} V^2)P^2K^2
\label{Eq:ModInst6}
\end{equation}
for the perturbation wave is obtained. Two characteristic scenarios for Eqs. (\ref{Eq:ModInst4}) and (\ref{Eq:ModInst5}) are possible based on Eq. (\ref{Eq:ModInst6}):
\begin{enumerate}
	\item if Q/P$<$0, all the possible values of $\Omega$ are real and perturbations only travel with a constant shape around the system. The system is therefore neutrally stable; 
	\item if Q/P$>$0,  $\Omega$ assumes imaginary values if $0<K<V\sqrt{2Q/P}$. In this case, perturbations grow exponentially in time and modulation instability  refereed to as self-modulation or side-band instability arises.  
\end{enumerate}

\begin{figure}[h]
	\centering
	\includegraphics[trim=4.9cm 0.25cm 4.5cm 0.05cm, clip=true, angle=0, scale=0.275]{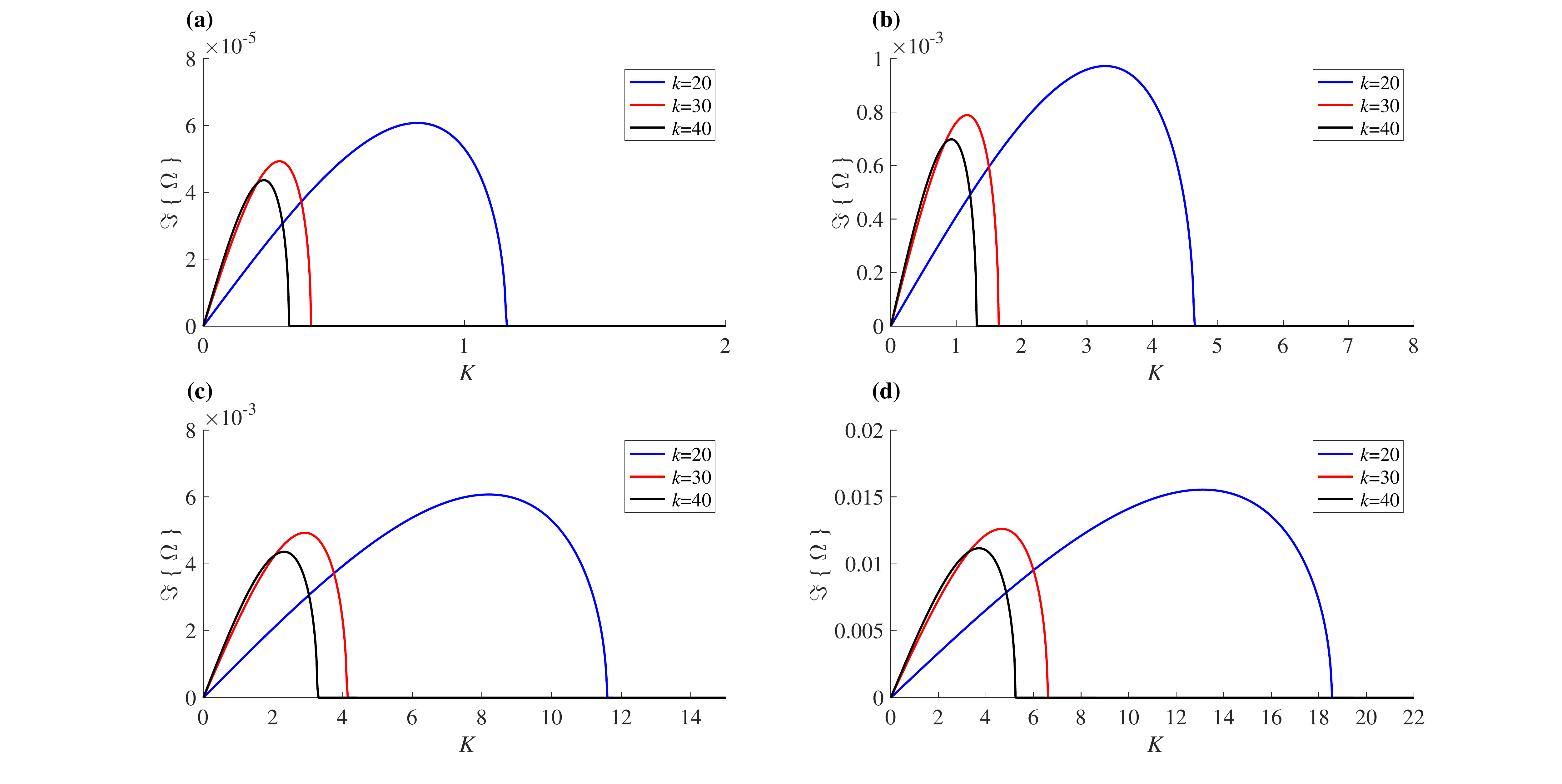}
	\caption{The imaginary part of $\Omega$ as a function of the carrier wave number for different values of $K$ and $V$. Panels (a), (b), (c), and (d) display the values of $\Im\{\Omega\}$ for $V=0.025 \ m$, $V=0.10 \ m$, $V=0.25 \ m$, and $V=0.40 \ m$, respectively. The curves of each panel represent the same quantities for different values of $k$.}  
	\label{Fig:Stab}
\end{figure}

Figure \ref{Fig:Stab} displays the instability diagram for different carrier wave numbers and amplitudes calculated from Eq. (\ref{Eq:ModInst6}) and based on the parameters assumed for the system, as described in Fig. \ref{Fig:PhySys}.
The obtained results state that the system is stable for almost all values of $k$ when the background amplitude is small and the system is almost in the linear regime. In the example in Panel (a) of Fig. \ref{Fig:Stab}, calculated by assuming $V=0.025 \ m$, the only unstable solution is obtained for $k=20$. All other curves predict instability for $K<1$, which would represent modes that do not verify the cyclic condition. The regime of instability is increased in Panels (b), (c), and (d) since the values of $V$ are bigger and the effects of nonlinearities are stronger.

\begin{figure}[h]
	\centering
	\includegraphics[trim=.075cm 0.05cm 0.5cm 0.05cm, clip=true, angle=0, scale=0.24]{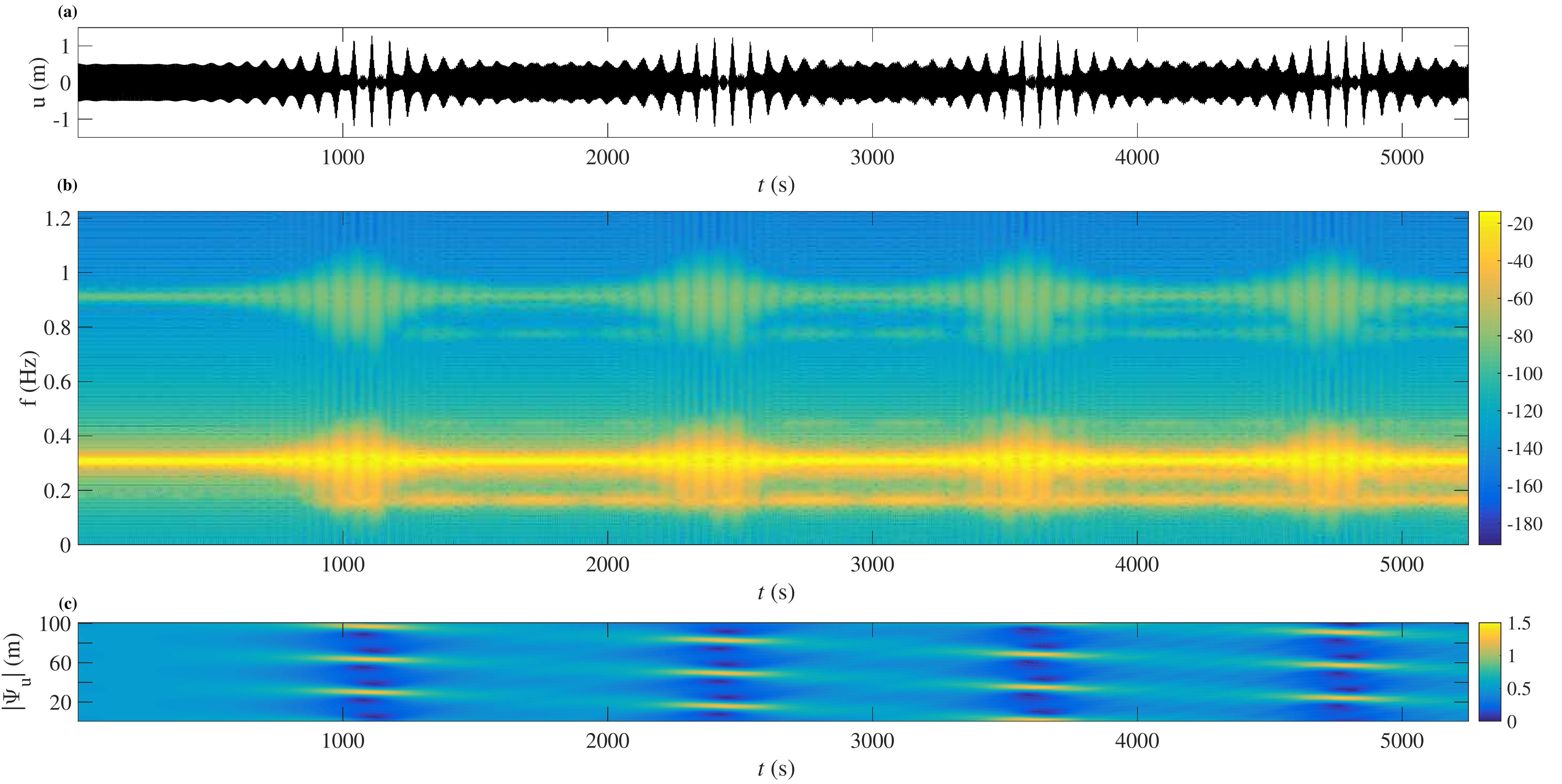}
	\caption{Stability of plane waves for $V=0.25 \ m$ and $k=30$ when a small perturbation with $K=3$ is added as an initial condition. Panel (a) displays the displacement of a specific degree of freedom, Panel (b) is the corresponding spectrum, in logarithmic scale, while Panel (c) illustrates the spatial envelope function in a frame moving with the group velocity.}
	\label{Fig:Spectrogram1}
\end{figure}

In order to compare the NLSE prediction with the full model system, a modulationally unstable plane wave is simulated and the corresponding results are shown in Fig. \ref{Fig:Spectrogram1}.   
The initial condition consists of a plane wave, with $V=0.25 \ m$ and $k=30$, onto which a small perturbation of wave number $K=3$ is added. This choice of parameters corresponds to the maximal growth rate from the NLSE theory (see Panel (c) of Fig. \ref{Fig:Stab}). The result in Panel (a) shows very interesting features. At first, the perturbation grows exponentially in time, as expected from the linear stability analysis. Subsequently, the solution reaches a maximum vibration amplitude and decays to a state very close to the initial plane wave condition. After reaching this configuration, recurrence is observed and an apparently regular vibration pattern  seems to be achieved. This type of dynamics can be attributed to the Fermi-Pasta-Ulam recurrence \cite{Fermi1955,Kimmoun2016}. The same response is investigated in the frequency domain, as plotted in Panel (b). The initial condition shows basically the main carrier wave frequency and its first harmonic ($f_2=3f_1 \sim 0.57$), which is an intrinsic nonlinear feature for the simulated plane wave. The initial perturbation grows and reaches the maximum amplification, as seen for $t \sim 1000$ seconds,  when the displacement spectrum becomes broader. When decaying to the initial configuration, the spectrum narrows again and looks very similar to the initial condition. The spatial configuration of the system and its evolution is investigated in Panel (c), where the envelope function $\Psi_u$, obtained from the Hilbert transform \cite{Feldman2011}, is displayed in a frame moving with the linear group velocity $c_g$. Firstly, it is possible to clarify that the system self-localises vibrations in three-humps equally spaced over its spatial domain in the co-moving frame. This pattern is usually referred to as breather dynamics in the literature, and it will be the focus of further investigation in the next subsection.

\subsection{Breathers}

\begin{figure}[h]
	\centering
	\includegraphics[trim=.25cm 0.75cm 1cm 0.25cm, clip=true, angle=0, scale=0.24]{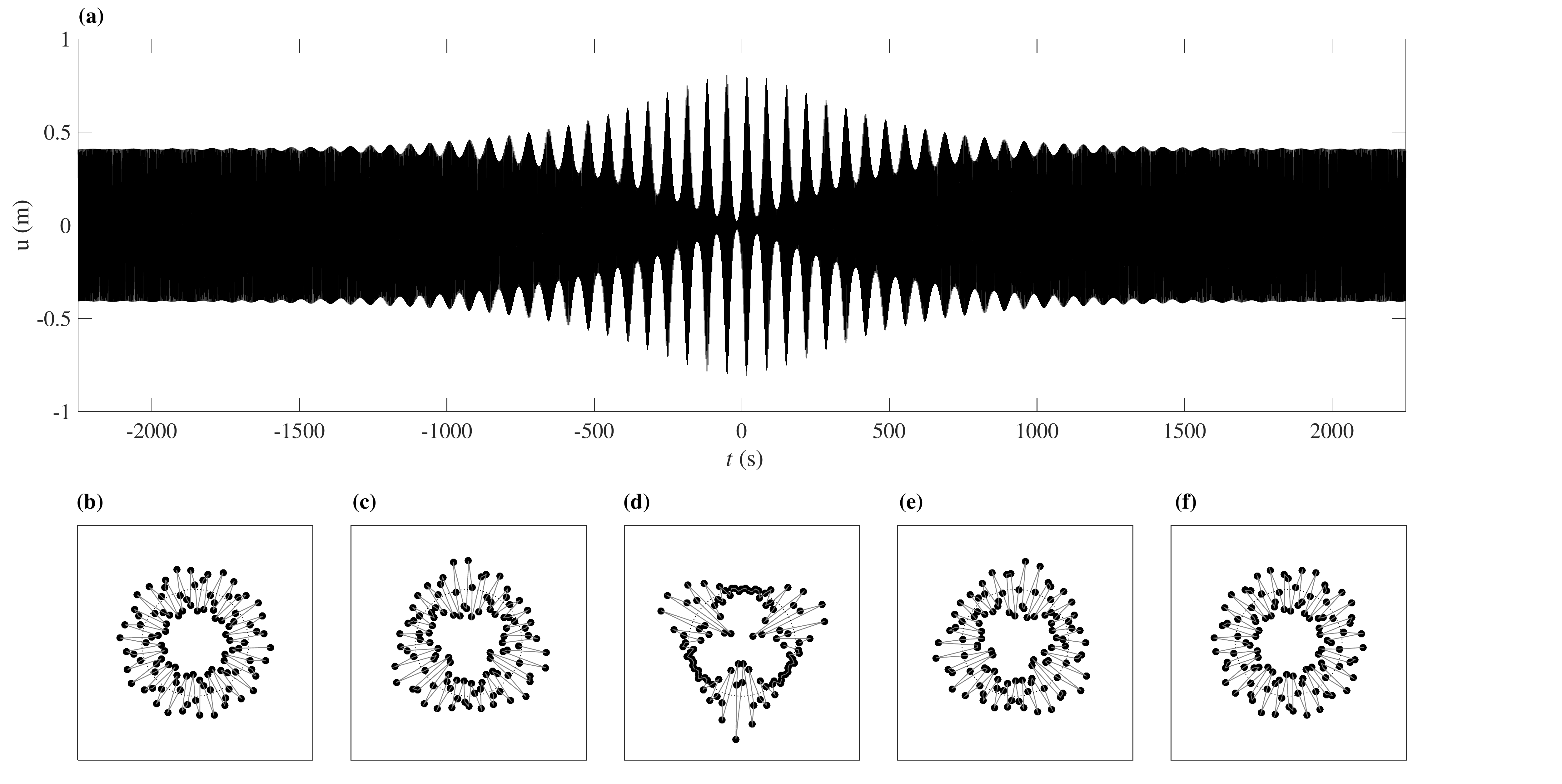}
	\caption{Akhmediev breather simulated in the cyclic symmetric system. Panel (a) shows a degree of freedom displacement at which maximum amplification is expected, while Panels (b), (c), (d), (e) and (f) illustrate the spatial configuration at $t=-2250 \ s$, $t=-500 \ s$, $t=0 \ s$, $t=500 \ s$, and $t=2250 \ s$, respectively.}
	\label{Fig:AKb}
\end{figure}

Breathers are a class of localised solution that has recently attracted considerable attention in the literature \cite{Onorato2013,Kibler2012}. They consist of exact analytic solutions that describe the evolution of modulationally unstable plane waves in the NLSE context. Nonlinear modulation dynamics in general deals with spatial, temporal, or spatio-temporal localisation during the time evolution. Before addressing specific solutions, the standard NLSE presented in Eq. (\ref{Eq:NLS}) is rewritten in another dimensionless form. This transformation is obtained by assuming that both $P$ and $Q$ have negative values, such that 
\begin{equation}
i \dfrac{\partial \Phi}{\partial \zeta} + \dfrac{1}{2} \dfrac{\partial^2 \Phi}{\partial \vartheta^2} + |\Phi|^2\Phi=0,
\label{Eq:DimNLS2}
\end{equation} 
where $\vartheta=\dfrac{\eta}{\sqrt{-2P}}$, $\zeta=-\tau$, and $\Psi(\tau,\eta)=\dfrac{\Phi(\zeta,\vartheta)}{\sqrt{-Q}}$. It is possible to show that Eq. (\ref{Eq:DimNLS2}) admits the following localised solution
\begin{eqnarray}
\Phi(\zeta,\vartheta)=V \left\lbrace 1 + \dfrac{2(1-2\alpha)\cosh(V^2b\zeta) + ib\sinh(V^2b\zeta)}{\sqrt{2a}\cos(V \kappa  \vartheta) - \cosh(V^2b\zeta)} \right\rbrace\nonumber
\times \exp\{iV^2\zeta\},
\label{Eq:AB}
\end{eqnarray}
first reported in Refs. \cite{akhmediev1985,Akhmediev1987}. In Eq. (\ref{Eq:AB}) the variable $V$ denotes the solution amplitude, while $b=\left[8\alpha(1-2\alpha)\right]^{1/2}$ and $\kappa=2(1-2\alpha)^{1/2}$ are two other constants. The solution dynamics is fully controlled by the free parameter $\alpha$, which basically sets the amount of energy localisation. In this case, if $\alpha<1/2$ the solution is known as an Akhmediev breather (see Panel (c) of Fig. \ref{Fig:Appendix}), while if $\alpha>1/2$ the solution is known as a Kuznetsov-Ma breather. Several configurations are possible for Eq. (\ref{Eq:AB}), depending on the number of solution humps and the maximum amplification. For the case of Akhmediev breathers a required configuration is implemented, considering the argument periodicity in the cosine function of Eq. (\ref{Eq:AB}), as  
\begin{equation}
V \kappa  \vartheta = \dfrac{2V(1-2\alpha)^{1/2}\eta}{\sqrt{-2P}}=n_h \ \ \textnormal{and} \ \ \alpha=\dfrac{1}{2} \left\lbrace 1 + \dfrac{Pn_h^2}{2V^2} \right\rbrace  ,
\end{equation} 
where $n_h$ represents the number of localised humps in the proposed solution. One should note that, due to the finite number of discrete oscillators, the solution does not have physical meaning for an arbitrary high number of localised humps. Figure \ref{Fig:AKb} 
illustrates the dynamics of a simulated result calculated by using $k=30$, $n_h=3$, and $\dfrac{V}{\sqrt{-Q}}\sim 0.2 \ m$. The initial condition is characterized by a carrier wave with an arbitrarily small perturbation. The perturbation grows over time and results, at $t\sim0 \ s$, in three very localised humps symmetrically spaced over the spatial domain. The observed amplification factor is around  1.98, which is in good agreement with the theoretical solution of Eq. (\ref{Eq:AB}), which predicts the amplification as 1.92 for the same system parameters. In addition, Panel (a) of  Fig. \ref{Fig:AKb} also indicates that the maximum amplification factor is not observed for $t=0 \ s$, but later. This feature indicates that the simulated results move more slowly than the linear group velocity. In fact, discrepancies like this are often observed when NLSE results, which are valid only for asymptotically weak nonlinearity, are compared to results obtained from full system models with weak, but non-zero nonlinearity \cite{Remoissenet1994}.

\begin{figure}[h]
	\centering
	\includegraphics[trim=.25cm 0.75cm 1cm 0.25cm, clip=true, angle=0, scale=0.24]{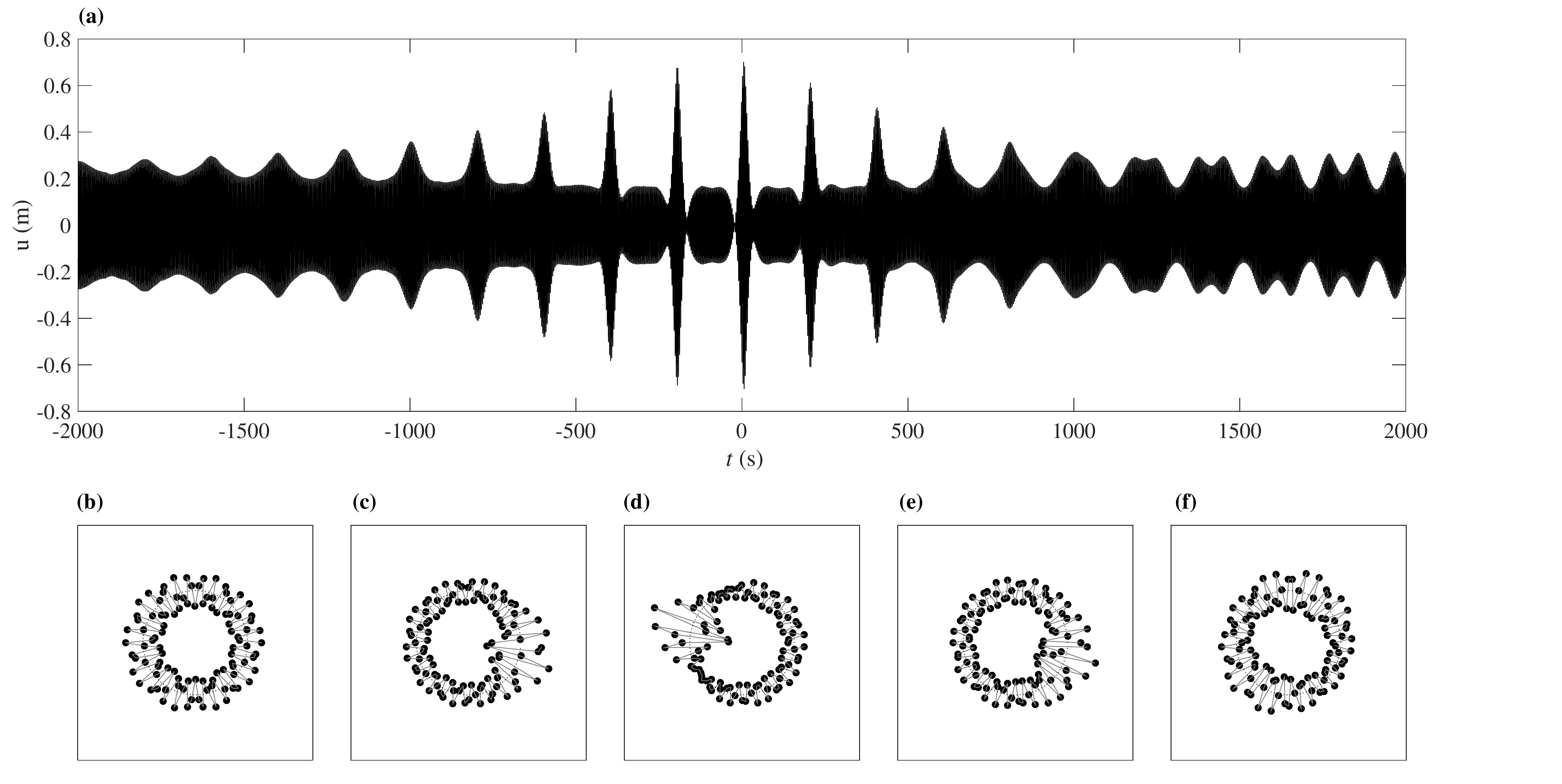}
	\caption{Peregrine breather simulated in the cyclic symmetric system. Panel (a) shows the temporal evolution for the mass at which maximum amplification factor is expected, while Panels (b), (c), (d), (e) and (f) depict the spatial configuration at $t=-2000 \ s$, $t=-500 \ s$, $t=0 \ s$, $t=500 \ s$, and $t=2000 \ s$, respectively. }
	\label{Fig:Pb}
\end{figure}

For the special case when $\alpha=1/2$, it is possible to write down a limiting case, known as a Peregrine breather, expressed by the rational doubly-localised solution \cite{Peregrine1983} such that
\begin{equation}
\Phi(\zeta,\vartheta)=V \left[ 1  - \dfrac{4(1+2iV^2\zeta)}{1 + 4 V^4 \zeta^2 + 4V^2 \vartheta^2} \right] \exp\{iV^2\zeta\},
\end{equation}
where $V$ is, again, the background amplitude. This equation has important features since it corresponds to the maximum localisation regime with an amplification factor of 3.00 at $\zeta=\vartheta=0$ (see Panel (c) of Fig. \ref{Fig:Appendix}). In addition, it consists of a hump that breathes only once in space and time, reflecting a single event that \textquotedblleft appears from nowhere and disappears without a trace\textquotedblright \cite{Akhmediev2009}. This solution has been proposed as a prototype for rogue waves, a class of very extreme and rare events well-known from optics and hydrodynamics and also recovered in laboratories \cite{Akhmediev2009,kibler2010,Chabchoub2011}. Figure \ref{Fig:Pb} displays the obtained results calculated by assuming $k=30$ and $\dfrac{V}{\sqrt{-Q}}\sim 0.12 \ m$. The maximum displacement consists of an amplification factor of 3.03, which is in very close agreement with the value 3.00 expected from the Peregrine solution. 

In the context of vibrations the Peregrine breather is a remarkable solution too. When starting with a plane wave in the focusing wave regime, an arbitrarily small perturbation, if chosen appropriately, can lead to substantial spatio-temporal localisation of vibration amplitude. For the inexperienced observer it might seem as if a travelling wave undergoes a spontaneous nonlinear distortion, at which a strong amplitude focusing with displacements three times higher than the original plane wave appears. Finally, the modulation disappears and the system experiences plane wave dynamics again. 

\begin{figure}[h]
	\centering
	\includegraphics[trim=.25cm 0.75cm 1cm 0.25cm, clip=true, angle=0, scale=0.24]{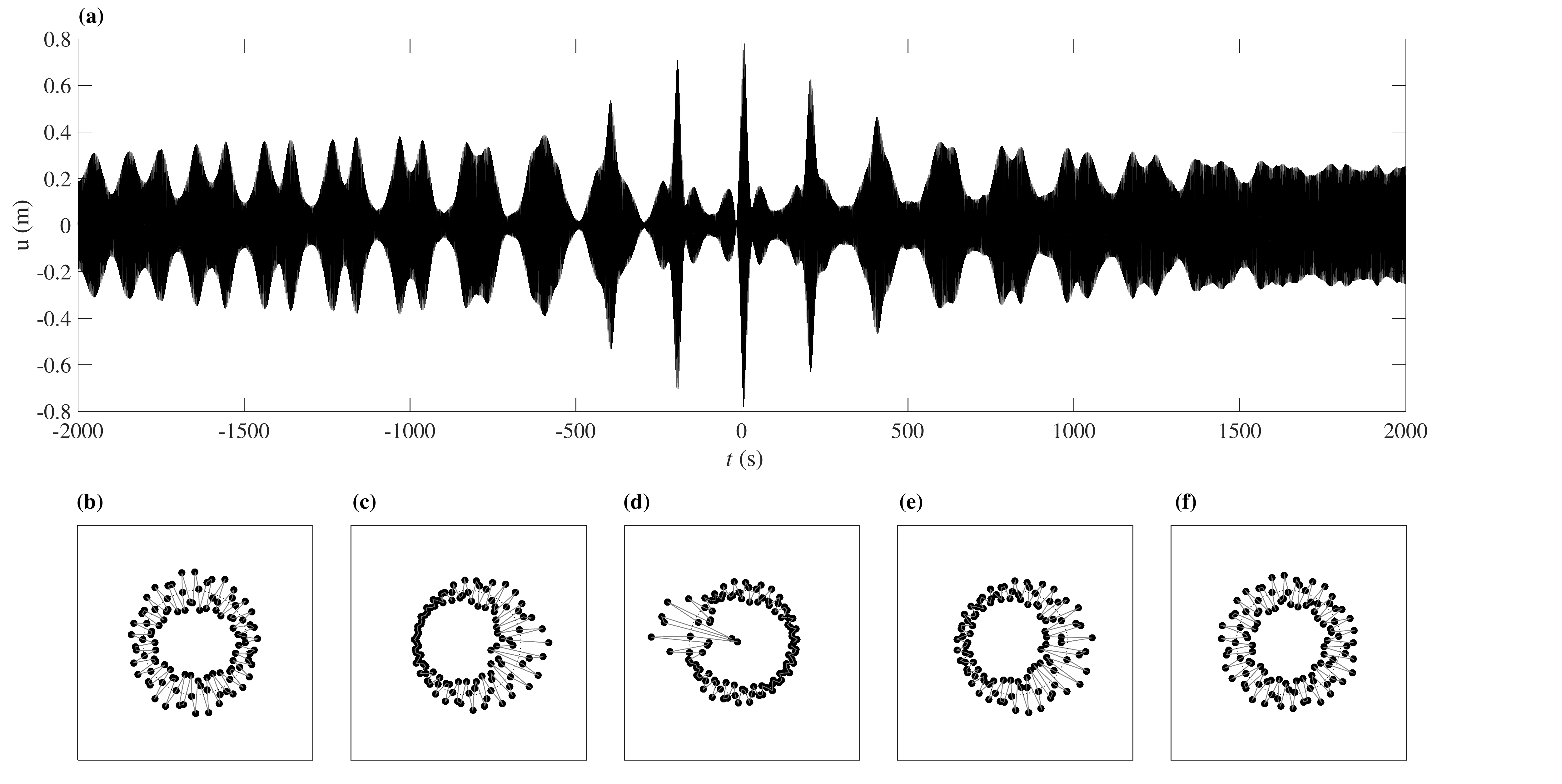}
	\caption{Akhmediev-Peregrine solution. The panels of this figure illustrate the same information of Fig. \ref{Fig:Pb}.}
	\label{Fig:HOb}
\end{figure}

Theory has long suggested that there are even more extreme nonlinear modulation effects, and solutions with higher localisation features have recently been observed e.g. in optics \cite{Erkintalo2011} and water waves \cite{Chabchoub2012,Chabchoub2012b}. The phenomenon basically consists of a nonlinear superposition of fundamental localised solutions, and analytical expressions are usually obtained by applying Darboux transform techniques. Solutions can be written down in a generic form as 
\begin{equation}
\Phi(\zeta,\vartheta)=V \left[ 1 - \dfrac{G(\zeta,\vartheta) - iH(\zeta,\vartheta)}{D(\zeta,\vartheta)}\right]  \exp\{iV^2\zeta\}, 
\label{Eq:HOB}
\end{equation}
where $G(\zeta,\vartheta)$, $H(\zeta,\vartheta)$ and $D(\zeta,\vartheta)$ are three functions stated in Refs. \cite{akhmediev1985,Akhmediev1986,Akhmediev2009} and they are presented in the Appendix. Equation (\ref{Eq:HOB}) represents an infinite hierarchy of localised solutions (see Panel (e) of Fig. \ref{Fig:Appendix}) which can, theoretically, represent unlimited amplification factors \cite{Akhmediev1991}. An example for a higher-order simulation result, known as the Akhmediev-Peregrine solution, is depicted in Fig. \ref{Fig:HOb}. In this case, the observed amplification factor is around 4.4, while the theoretical one is 5.0 for the corresponding NLSE solution. The discrepancies are probably due to higher-order nonlinearities and discreteness effects that are not considered in the NLSE approximation. Simulations have been calculated by assuming $\dfrac{V}{\sqrt{-Q}}\sim 0.09 \ m$ and $k=30$.
	
\section{Evolution of random initial conditions} \label{Sec:RandSol}

The evolution of random initial conditions can be regarded as highly relevant for practical applications in vibrations, where initial conditions come as a result of a multitude of uncontrolled factors. In the following we thus report simulation results for our model system subject to random initial conditions. The study has also been motivated by several recent similar studies in optics \cite{Walczak2015,Dudley2014}, where authors have reported very strong self-modulation when random initial conditions are considered. 

All simulated random initial conditions $u^{rand}_n(t=0)$ consist of plane waves $u^{PW}_n(t=0)$ on which random perturbation are added such that 
\begin{equation}
u^{rand}_n(t=0)= u^{PW}_n(t=0) + \sigma N_n,
\label{Eq:RandInitCond}
\end{equation}
where $N_n$ is a Gaussian random variable with zero mean and unit standard deviation. The noise level is fully controlled by the free parameter $\sigma$, which sets the final standard deviation of $\sigma N_n$. In order to obtain a very broad-band noise spectrum, the random signals $N_n$ and $N_{n\pm1}$ are taken as uncorrelated for all the following investigations.

\begin{figure}[h]
	\centering
	\includegraphics[trim=.075cm 0.05cm 0.5cm 0.05cm, clip=true, angle=0, scale=0.24]{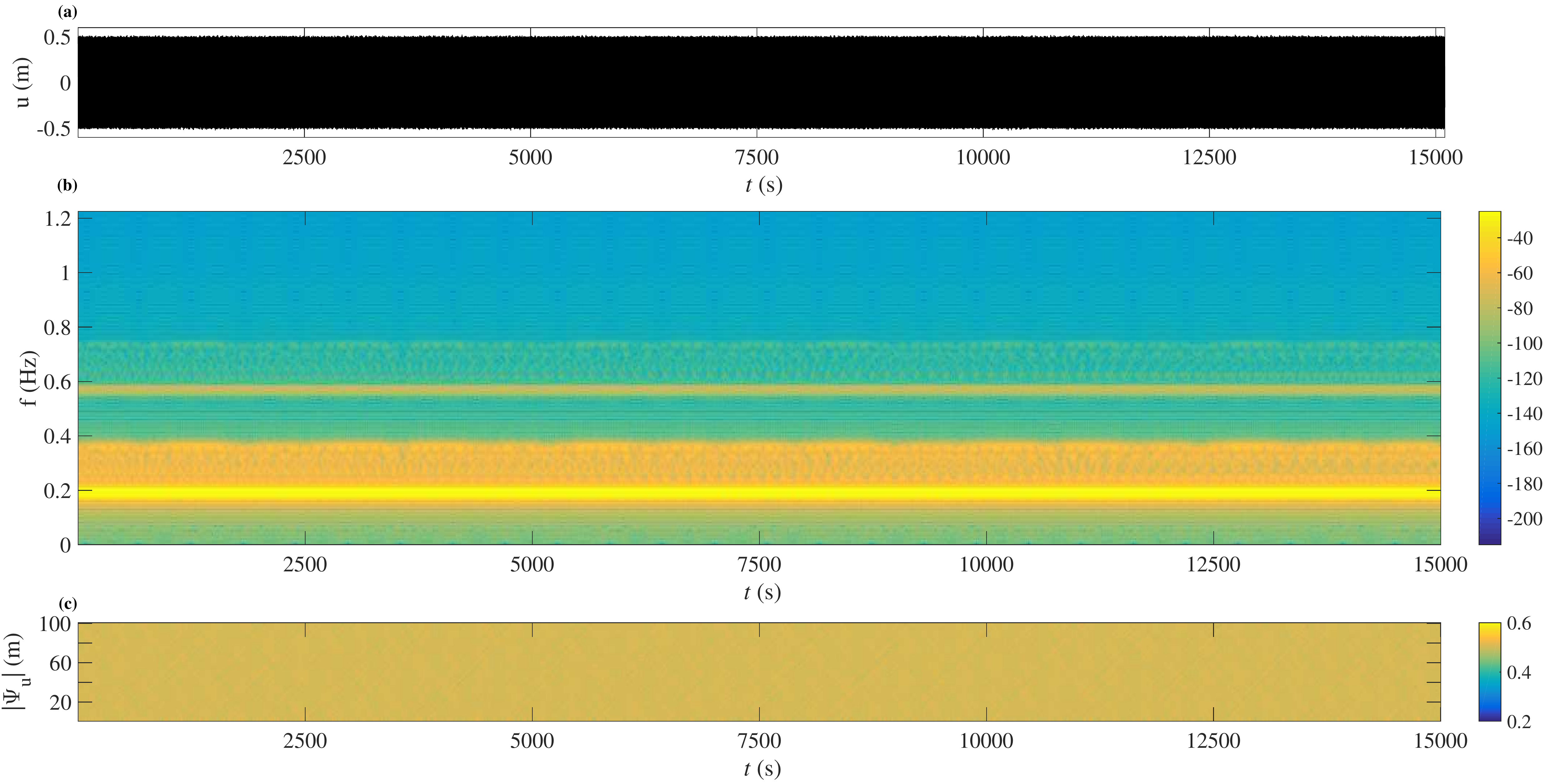}
	\caption{Stability of plane waves for $V=0.25 \ m$ and $k=10$ when a small Gaussian noise of zero mean and standard deviation $\sigma=0.01$ is added as an initial condition. Panels (a), (b) and (c) illustrates the same quantities of Fig. \ref{Fig:Spectrogram1}.}
	\label{Fig:Spectrogram2}
\end{figure}

Firstly, a perturbed plane wave is simulated in the defocusing range. In this scenario, the system is known to be linearly stable with respect to all perturbation wave numbers. The respective results, calculated by assuming plane wave parameters $V=0.25$ and $k=10$, and noise level $\sigma=0.01$, are displayed in Fig. \ref{Fig:Spectrogram2}. 
The displacement of a specific degree of freedom, depicted in Panel (a), shows that the small noise does not grow or decay significantly over time. This conclusion is confirmed looking at the results in the frequency range, as illustrated on the plot in Panel (b). It is shown that the spectrum consists of a main signal corresponding to the plane wave, at $f_1\sim0.19$ Hz, and a weaker response bounded between [$\omega_0^2$,$\sqrt{\omega_0^2+4\omega_c^2}$]. The initial pattern does not change significantly over the simulation. The envelope function, in Panel (c), shows that no localisation mechanism is observed in the spatial configuration while the system evolves in time.   

A very similar investigation is performed in the focusing range by using the same parameters as in Fig. \ref{Fig:Spectrogram2}, except for the wave number which is assumed as $k=30$.
\begin{figure}[h]
	\centering
	\includegraphics[trim=.075cm 0.05cm 0.5cm 0.05cm, clip=true, angle=0, scale=0.24]{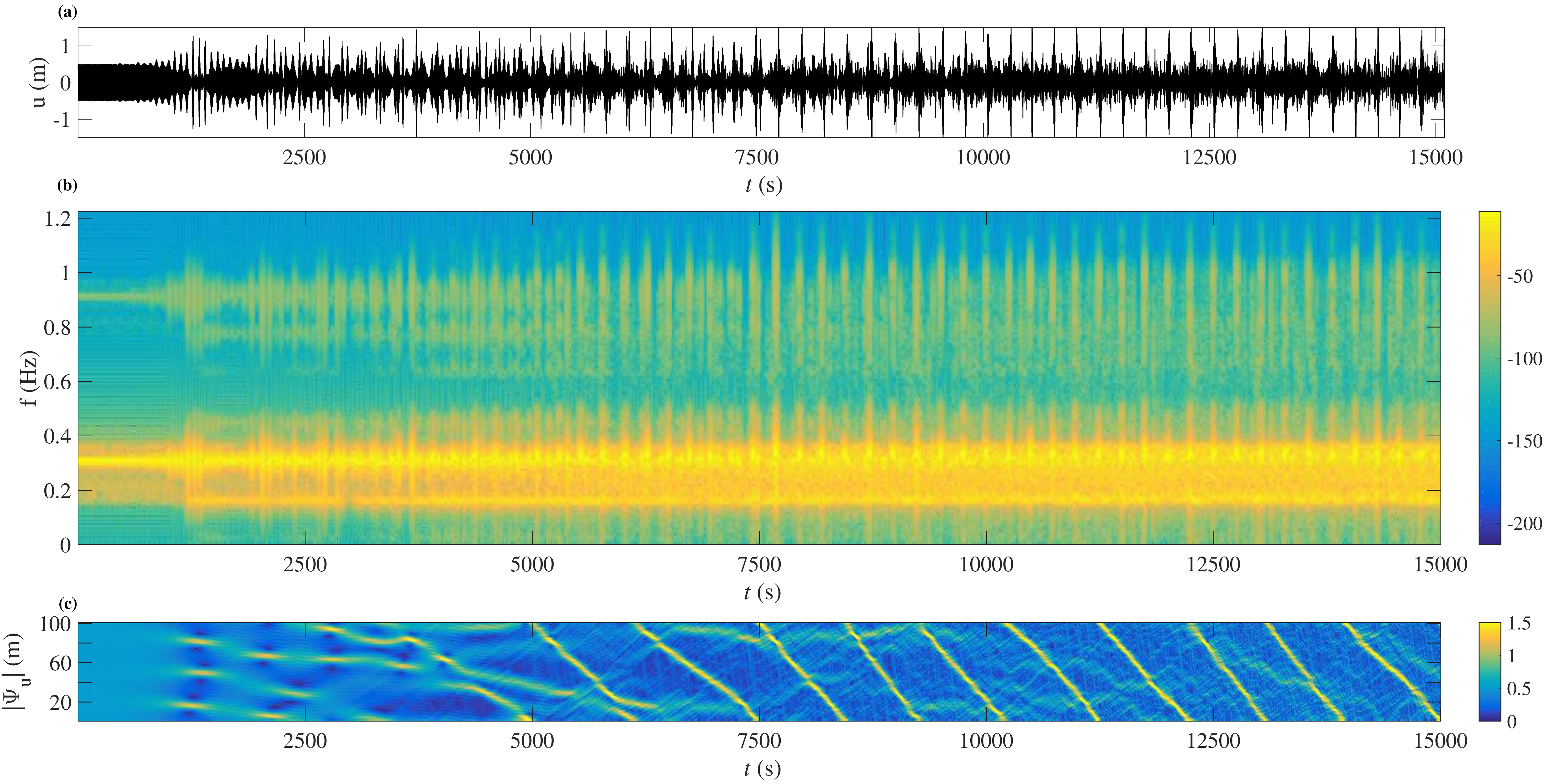}
	\caption{Stability of plane waves for $V=0.25 \ m$ and $k=30$ when a small Gaussian noise of zero mean and standard deviation $\sigma=0.01$ is added as an initial condition. Panels (a), (b) and (c) illustrate the same quantities of Fig. \ref{Fig:Spectrogram1}.}
	\label{Fig:Spectrogram3}
\end{figure}
Figure \ref{Fig:Spectrogram3} depicts the obtained results. From Panel (a) it is observed that the displacement at first consists of an Akhmediev breather-like solution that breathes only once before the system goes to an irregular regime. In the frequency domain, shown in Panel (b), the results are also interesting. The initial condition grows in a similar way to Akhmediev breathers, as seen for t $\sim$ 1500 $s$, although energy is transferred from the main signal to reach an almost flat response between [$\omega_0^2$,$\sqrt{\omega_0^2+4\omega_c^2}$] at the end of simulations. The spatial configuration is shown in Panel (c). Again, it is possible to confirm that the Akhmediev breather consists of three localised humps equally spaced over the system. The single and isolated perturbation wave number $K=3$ dominates the system dynamics and leads, as expected, to an exponentially growing modulation, followed by a dynamics very similar to the corresponding Akhmediev breather. Moreover, the transition from a breather response to an irregular one shows strong localised vibrations which consist of wave-packets travelling around the system with a certain constant velocity. Reconsidering the soliton solutions discussed earlier (see Ref. \cite{Grolet2016}), it seems that the breather-type dynamics has given way to a dynamics consisting mainly of envelope solitons. The latter is sometimes called soliton chaos or soliton turbulence \cite{Soto-Crespo2016}.

It should be noted that the observed transition from breather dynamics to soliton solutions has been thought to be unusual from a NLSE perspective. The NLSE is an integrable equation, and many insights for its solution are based on the Inverse Scattering Technique (IST). One of the major contributions, obtained from the IST, is that breathers and solitons might be seen as nonlinear modes of the problem and no transition from breathers to solitons would be expected within the NLSE context \cite{Akhmediev2016,Soto-Crespo2016}. However, a transition might be observed when higher-order effects are included within the NLSE approximation \cite{Mahnke2012}. For example, the NLSE approach is a very useful framework for soliton and breather turbulence in optics \cite{Walczak2015}, but it does not consider Raman scattering. In order to take this effect into account, a modified NLSE is required \cite{dysthe1979,Blow1989}. For such cases it has been observed that a transition from breather to soliton turbulence is possible \cite{Mahnke2012}. The time-integration results, obtained from the proposed physical system, seem to show similar features.

\section{Conclusions and outlook} \label{Sec:Con}

This paper explores vibration localisation due to weak nonlinearity in a chain of oscillators that allow dispersive travelling waves. The study focusses on the evolution of modulated travelling waves. The system under study may be seen as a minimal model for different aerospace structures, and consists of a cyclically connected chain of Duffing oscillators. In the defocusing range, dark solitons, characterized by a drop in modulation amplitude that moves around the system with a constant shape profile, do exist. In the focusing range, breather dynamics are computed and it is observed that they may be excited spontaneously when a travelling wave is slightly perturbed by random initial conditions. Finally, a transition from regular breather to irregular soliton dynamics is also observed. The obtained results suggest that very strong vibration localisation in cyclic structures might arise due to modulation instability and subsequent nonlinear evolution, potentially also including irregular soliton dynamics.  

Our studies have been based on a combination of results for the NLSE and numerical simulation for our model system. Future investigations will first need to consider the effects of damping and external forcing. More work also needs to be done on modelling. The results presented in this paper are based on a highly idealised minimal model, and obviously models capturing more realistic geometry and system properties, e.g. based on finite element models, will be in the focus of future work. For this case, the development of a NLSE based approach might be impracticable, and so efficient numerical procedures are required.

\section*{Acknowledgements}
	The first author thanks the financial support provided by the Brazilian agency CNPq (project number 01339/2015-3).

\bibliography{mybibfile3}

\appendix \section{Higher-order solutions}
The rational form of the Akhmediev-Peregrine solution investigated in this paper is written, assuming that $V$ is real, as follows \cite{Akhmediev1986,Akhmediev2009}:
\begin{eqnarray}
G(\zeta,\vartheta)=(V^2 \vartheta^2 + 4V^4\zeta^2 + \dfrac{3}{4})(V^2\vartheta^2 + 20V^4\zeta^2 + \dfrac{3}{4}) - \dfrac{3}{4},\\
H(\zeta,\vartheta)=2V^2\zeta(4V^4\zeta^2 - 3V^2\vartheta^2) + 2V^2\zeta\left[ 2(V^2\vartheta^2 + 4V^4\zeta^2)^2 -\dfrac{15}{8}\right], \\
D(\zeta,\vartheta)=\dfrac{1}{3} (V^2\vartheta^2 + 4V^4\zeta^2)^3 + \dfrac{1}{4}(V^2\vartheta^2 - 12V^4\zeta^2)^2 \nonumber\\ 
 +  \dfrac{3}{64}(12V^2\vartheta^2 +176V^4\zeta^2 + 1).
\end{eqnarray}

\end{document}